\documentclass[draftcls, onecolumn, 12pt]{IEEEtran}

\usepackage{cite}
\usepackage{amsmath,amssymb,amsfonts}
\usepackage{algorithmic}
\usepackage{graphicx}
\usepackage{textcomp}
\def\BibTeX{{\rm B\kern-.05em{\sc i\kern-.025em b}\kern-.08em
		T\kern-.1667em\lower.7ex\hbox{E}\kern-.125emX}}

\usepackage{bm}			
\usepackage{caption}
\usepackage{hyperref}

\usepackage{tikz}
\definecolor{gold}{rgb}{0.85,.66,0}
\definecolor{green2}{rgb}{0.01,0.5,0.01}
\definecolor{cian}{rgb}{.02,.7,.95}
\definecolor{ppp}{rgb}{.7,.3,.82}

\usepackage{cite}
\usepackage{color}
\usepackage{amssymb}
\usepackage{amsthm}

\begin{document}
\title{3-D Localization with Multiple LEDs Lamps in OFDM-VLC system}
\author{Luis C. Mathias, Leonimer F. de Melo,  Taufik Abr\~ao,
\IEEEmembership{Senior Member, IEEE}
\thanks{Department of Electrical Engineering, State University of Londrina, Rod. Celso Garcia, PR-445 Km 380, CEP 86051-990, Londrina-PR, Brazil  (e-mail: luis.mathias@uel.br; \quad leonimer@uel.br \quad taufik@uel.br)}
\thanks{``This work was supported in part by the  National Council for Scientific and Technological Development (CNPq) of Brazil under Grants 304066/2015-0; by the Londrina State University (UEL) and the Paran\'a State Government''}}
	
\markboth
{L. C. Mathias, \, L. F. Melo, \,  T. Abrao: 3-D Localization with Multiple LEDs Lamps in OFDM-VLC system}
	{L. C. Mathias, \, L. F. Melo, \, T. Abrao: 3-D Localization with Multiple LEDs Lamps in OFDM-VLC system}

\maketitle
	
\begin{abstract}
Visible light communication (VLC) based localization is a potential candidate for wide range indoor localization applications. In this paper, we propose a VLC architecture {based on} orthogonal frequency division multiplexing (OFDM) {with multiple functionalities  integrated in the same system, i.e., the 3-D receiver location, the control of the room illumination intensity, as well as the data transmission capability}.  Herein we propose an original methodology for LED power discrimination applying spatial optical OFDM (SO-OFDM) structure for position estimation. The hybrid locator initially makes a first estimate using a weighted angle-of-arrival (WAoA)-based locator which is then used as the starting point of the recursive estimator based on the strength of the received signal (RSS). Hence, the first stage is deployed to increase convergence probability, reducing the root-mean-square error (RMSE) and the number of iterations of the second stage. Also, a performance {\it vs} computational complexity comparative analysis is carried out with parameter variations of these estimators. The numerical results indicate a decade improvement in the RMSE for each two decades of decrement of power noise on the receiver photodiode.  The best clipping factor is obtained through the analysis of locator accuracy and transmission capacity for each simulated system. Finally, the numerical results also demonstrate effectiveness, robustness, and efficiency of the proposed architecture.
\end{abstract}
\begin{IEEEkeywords}
 3-D Position estimation, AoA, OFDM, RSS, VLC.
\end{IEEEkeywords}
	
\section{Introduction}
\label{sec:introduction}
Visible Light Communication (VLC) concept has now gained prominence due to the availability of a vast and still unexplored spectral band in the frequency range of visible light, aiming at facing with the growing demand for data transmission. VLC provides exceptionally high transmission rates to the end user in a scenario of increasing frequency spectrum shortage in RF communication systems. 

The research related to a 3-D location in VLC environment has been promising due to several factors. The first occurs in applications where the Global Positioning System (GPS) signal cannot penetrate the environment application. The second is due to the increasing replacement of conventional lamps with those of light emitting diode (LED) that are more long-lasting and of better energetic and luminous efficiency. In this case, the LED lamp infrastructure in addition to illuminating, can transmit data and can also allow the localization of a mobile receiver. Thus, for a practical VLC system, it is desirable to use the same transmission technology for both positioning and high-speed data transmission. 

Due to the low cost and low complexity, the Intensity Modulation with Direct Detection (IM/DD) is the most practical method of implementing a VLC system. In this modulation type, the electric current of the LED transmitters is modulated to vary the transmitted light intensity. At the receiver side, the received light intensity is converted directly into electrical current utilizing photodetectors. Thus, in the IM/DD, it is necessary that the signal in time must be real and positive so that the light intensities of the LED transmitters are modulated directly \cite{2012_ghassemlooy}. 

A baseband modulation technique that is extensively exploited in IM/DD due to the efficient use of the available bandwidth is the Orthogonal Frequency Division Multiplexing (OFDM) \cite{ofdm_owc}. Since, usually, a time OFDM signal is bipolar, several modulation techniques are found in the literature to make it unipolar \cite{dco_aco_flip_ofdm}. In this way, a real OFDM time-frame is achieved by imposing a Hermitian symmetry on the vector of symbols mapped previous the Inverse Fast Fourier Transform (IFFT) block. In DC-biased optical OFDM (DCO-OFDM), a bias signal is added to make the time-signal positive. On the other hand, in the asymmetrically clipped optical OFDM (ACO-OFDM), the transmitted signal is produced positive by sending only the odd subcarriers. The Flip-OFDM divides the original OFDM frame into two parts by transmitting them separately \cite{flip_ofdm}. The first frame is reassembled with positive points in time, and another frame is formed by inverting the polarity of the points in time that were negative. The ACO-OFDM and Flip-OFDM are commensurate regarding spectral efficiency and error performance, but the Flip-OFDM save nearly half of receiver complexity over ACO-OFDM. However, the two techniques have approximately half the spectral efficiency compared to the DCO-OFDM \cite{2017khalighi}. 

Two techniques are usually used in estimating the location of a VLC receiver. The first one, the angle of arrival (AoA), takes the direction of the LED transmitters into consideration at the receiver side. The second technique, based on the location by Received Signal Strength (RSS), considers the strength of the signal captured by the receiver due to the transmitter LEDs \cite{2016_3-D_RSS}.  In \cite{2014_AOA_RSS_KALMAN} also is proposed an integrated AoA-RSS localization method that finds out the 2-D position of a mobile robot using an array of photodiodes (PDs). Moreover, the work \cite{IEEEhowto:sahin} deals with the 3-D localization problem and uses these two localization techniques. Although the location by RSS is more accurate than the AoA-based method, in general, its recursive estimator presents a non-convex structure and can achieve different results than those expected \cite{IEEEhowto:sahin}. In this way, the position estimation obtained by the AoA locator can be used as the initial search point for the RSS locator. Such strategy has the purpose of starting the search at a closer location, reducing the possibilities of divergence. This hybrid estimator considers the RSS information of each LED separately. It is worth to note that in the previous works, such as \cite{2014_AOA_RSS_KALMAN} and \cite{IEEEhowto:sahin}, no scheme has proposed for the discrimination of the light powers received from each LED. 

The works on the OFDM transmission scheme for the localization estimation purpose present just only 2-D estimators. For instance, the work \cite{2016_OFDMposition} employs {training symbol} in OFDM and uses RSS information for estimates of the distances between the receiver and the LED transmitter by applying a {\it lateration} technique. Such technique handles the geometric analysis of the problem through triangles and circles. Moreover, the work \cite{experimental_ofdma_locator} reports an experimental demonstration of an indoor 2-D VLC positioning system based on the OFDM transmission scheme that proposes to discriminate the signals transmitted by three different LEDs using coding in three OFDM subcarriers. Thus, the receiver retrieves all signals transmitted using a Discrete Fourier Transform (DFT) operation.  A Spatial Optical-OFDM (SO-OFDM) scheme with multiple LEDs is proposed in \cite{vlc_multiple_leds} trying to mitigate the OFDM Peak-to-Average Power Ratio (PAPR) problem in VLC. In this design, filtered subsets of OFDM subcarriers are emitted by each LED,  allowing the receiver to discriminate the power received from each LED.

The {\it contribution} of this work is threefold, as summarised in the following. {\bf a}) It is proposed a VLC structure with power discrimination at the receiver side aiming at improving the 3-D indoor localization feature;  {\bf b})  an innovative hybrid 3-D building localization scheme is proposed that distributes a {training symbol} for each LED among the subcarriers arranged into SO-OFDM groups;  {\bf c}) based on extensive numerical simulations results, our hybrid location estimator can be implemented allowing a more precise 3-D location by considering more RSS information of each LED from multiple LED lamps infrastructure. 

The paper is divided into five sections. Besides this introductory section, the {Section} \ref{sec:hybrid} develops the VLC system model deployed in the AoA and RSS estimators which are described in {Subsections} \ref{subsec:rss} and \ref{subsec:aoa} respectively. In Section \ref{sec:multiple_leds}, the 3-D hybrid estimator obtained is applied to the SO-OFDM multiplexing scheme with DCO-OFDM. In Section \ref{sec:resultados} numerical simulation results are considered aiming at corroborating the quality of the 3-D location estimations for the proposed scheme. Finally, in Section \ref{sec:conclusion} the conclusions are offered.

{ \noindent Notation: $\bm{I}_K$ is the $K \times K$ identity matrix and the field of real numbers is denoted by $\mathbb{R}$. The transpose  operation and the Moore-Penrose are denoted by $(.)^\intercal$ and $(.)^\dagger$ respectively. The matrix kernel is denoted by $ker(.)$.  
	$\mathcal{N}(0,\bm{C} )$ means the Gaussian distribution with zero mean and covariance matrix $\bm{C}$. $\mathcal{U}(a,b)$ holds for a uniform distribution with boundaries $a$ and $b$.}

\section{Hybrid Localization}\label{sec:hybrid}
This section describes the system and the noise models, as well as the RSS and the AoA position estimators. 

\subsection{VLC System Model}\label{subsec:system}
\begin{figure}[!htbp]
	\centering
	\includegraphics[width=.45\textwidth]{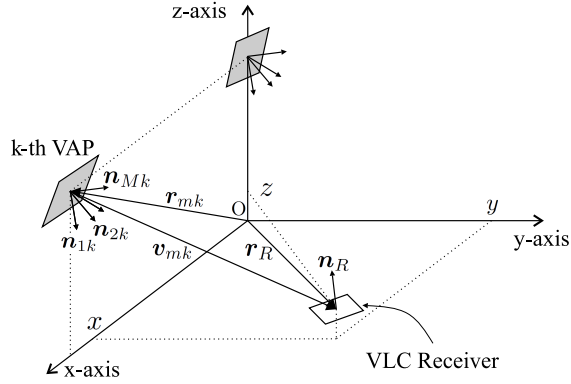}
	\caption{Schematic diagram representation for the 3-D localization problem in a VLC system \cite{IEEEhowto:sahin}.}
	\label{fig:system_model}
\end{figure}

The VLC system can be modeled considering $K$ visible light access points (VAP) with some $M$ elements LEDs transmitting each \cite{IEEEhowto:sahin}.  Fig. \ref{fig:system_model} presents a schematic diagram of the 3-D system model depicting the position of vectors, normal versors, and angles in a Cartesian plane.

\begin{figure}[!htbp]
	\centering
	\includegraphics[width=.65\textwidth]{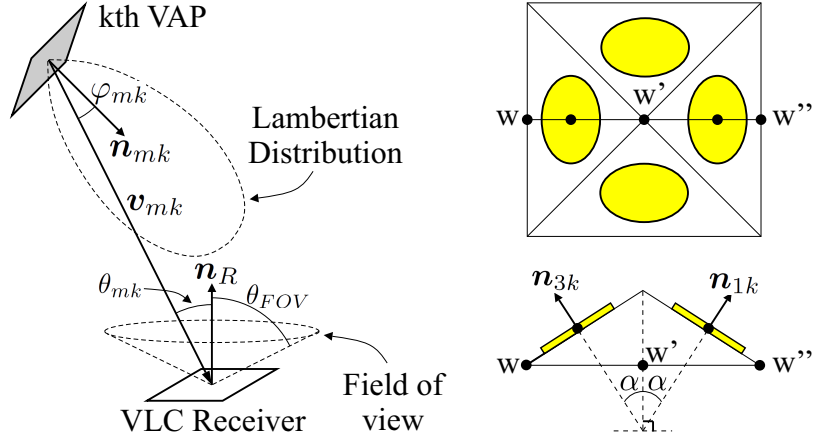}
	\caption{Details of the angles involved in the model and example of a VAP with 4 LEDs in pyramidal format.}
	\label{fig:system_model2}
\end{figure}

The position vectors and the orientation versor of the receiver is denoted by $\bm{r}_R  = [x,y,z]^\intercal$  and $ \bm{n}_R =[n_R ^{(x)} ,n_R ^{(y)},n_R ^{(z)}]^\intercal$. The position vectors and the orientation versor of the $m$-th LED transmitter of the $k$-th VAP are $\bm{r}_{mk}  = [x_{mk} ,y_{mk} ,z_{mk} ]^\intercal $ and $ \bm{n}_{mk}  = [n_{mk} ^{(x)} ,n_{mk} ^{(y)} ,n_{mk} ^{(z)} ]^\intercal$. Thus, the vector denoting the distance between the transmitting element and the receiver can be given by:    
\begin{equation}
\bm{v}_{mk}  = \bm{r}_R  - \bm{r}_{mk}  = [a_{mk} ,b_{mk} ,c_{mk} ]^\intercal  \in \mathbb{R}^{3 \times 1}.
\label{eq:vmk}
\end{equation}
Hence, the DC optical channel gain between the receiver and the $m$-th LED of the $k$-th VAP can be given by \cite{IEEEhowto:sahin}:
\begin{equation}
\Omega_{mk}  =  \kappa \cdot \prod \left( {\frac{{\theta _{mk} }}{{\theta _{FoV} }}} \right) \cdot \prod \left( {\frac{{\varphi _{mk} }}{{\pi /2}}} \right) \cdot f(\bm{v}_{mk}),
\label{eq:pot2}
\end{equation}
where
\begin{equation}
\kappa = - \frac{{(n_L + 1)A_{pd} }}{{2\pi }}
\end{equation}
and $f(\bm{v}_{mk})=$
\begin{equation}
\begin{array}{l}
\displaystyle
\frac{{\left( {\bm{v}_{mk} ^\intercal \bm{n}_{mk} } \right)^{n_L} \bm{v}_{mk} ^\intercal \bm{n}_R }}{{\left\| {\bm{v}_{mk} } \right\|_2^{n_L + 3} }} =  \left( {a_{mk} n_{mk} ^{(x)}  + b_{mk} n_{mk} ^{(y)}  + c_{mk} n_{mk} ^{(z)} } \right)^{n_L} \frac{{\left( {a_{mk} n_R ^{(x)}  + b_{mk} n_R ^{(y)}  + c_{mk} n_R ^{(z)} } \right)}}{{\left( {a_{mk} ^2  + b_{mk} ^2  + c_{mk} ^2 } \right)^{\frac{{n_L + 3}}{2}} }}, \\ 
\end{array}
\label{eq:fvmk}
\end{equation}
\normalsize
\noindent where $\varphi_{mk}$ is the angle between the orientation versor of the LED transmitter and the incidence vector, $\theta_{mk}$ is the angle between the receiver orientation versor and the incidence vector, $A_{pd}$ is the area of the photodetector (PD) in $m^2$, $\theta_{FoV}$ is the field of view (FoV) of the PD, $n_L$ is the mode number of the Lambertian distribution\footnote{{The Lambertian distribution is commonly used to describe the luminous distribution of LEDs. The distribution is more directive for higher values of $n_L$.}}. The FoV of receiver effect and the field of emission effect of LED transmitter are also considered in \eqref{eq:pot2} using the rectangular function defined by:
\begin{equation}
\prod (\cdot) \buildrel \Delta \over = \left\{ {\begin{array}{*{20}c}
	{1,} & {\left| \cdot \right| \le 1}  \\
	{0,} & {\left| \cdot \right| > 1}.  \\
	\end{array}} \right.
\end{equation}
Details of the transmission angle, incident angle, FoV, and example of a VAP arrangement with 4 LEDs in pyramidal format are presented in Fig. \ref{fig:system_model2}. A greater FoV is attractive because the location estimators can evaluate all the LED transmitting powers. In contrast, it exposes the receiver to a higher incidence of noise and interference.

{Assuming} that the transmitted optical power of each LED is equal to $P_T$, the optical power of the $m$-th LED of the $k$-th VAP in the receiver can be given by:
\begin{equation}
P_{mk}  =\Omega_{mk}  P_T.
\label{eq:power_light_received}
\end{equation}
{where $\Omega_{mk}$ is given by \eqref{eq:pot2}.} Therefore, the total power received by the photodetector in the VLC receiver is the sum of the optical powers received from each transmitter LED, i.e. $P_R  = \sum\nolimits_{m = 1}^M {\sum\nolimits_{k = 1}^K {P_{mk} } }$. Thus, the current generated is proportional to the power received with additive white Gaussian noise (AWGN) \cite{vlc_multiple_leds}. In this context, the electric gain $G_E$ can be given by:
\begin{equation}
G_E={S_{\textsc{led}} \Omega_{mk} R_{pd}},
\label{eq:electric_gain}
\end{equation}
where $S_{\textsc{led}}$ is the LED conversion factor in $[W/A]$ and $R_{pd}$ is the photodetector responsivity in $[A/W]$. Both parameters consider radiometric light power.

The photodetector responsivity $R_{pd}$ is generally presented in the datasheet of the PIN junction photodiodes. The LED conversion factor $S_{\textsc{led}}$ is a parameter that varies due to the non-linearity of the luminous flux $\phi_V$ in $[\rm{lm}]$ as a function of the electrical current $I_{\textsc{led}}$ in $[\rm{A}]$ \cite{2016chapterModelingLED}. This relation for the Cree\textsuperscript{\textregistered} XHP70.2 6V LED device can be modelled by a polynomial quadratic function as \cite{datasheetCreeLED}:
\begin{equation}
\phi_V(I_{\textsc{led}})=-31.29I_{\textsc{led}}^2 + 705.35I_{\textsc{led}} + 20.7.
\label{eq:phi_v}
\end{equation}

The conversion from luminous flux to the radiated optic power $P_T$ can be realized by a factor of $2.1[\rm{mW/lm}]$ for phosphor-coated blue LED \cite{2008broadbandLED}. Moreover, using predistortion with  upper and lower current limits of modulation $I_u$ and $I_l$, respectively, the LED conversion factor can be determined by \cite{vlc_multiple_leds}:
\begin{equation}
S_{{\textsc{led}}}= 0.0021\frac{{\left( {\varphi _V (I_u ) - \varphi _V (I_l )} \right)}}{{I_u  - I_l }}.
\label{eq:led_factor}
\end{equation}
\subsection{Noise Model}\label{subsec:noise}
The noise directly affects the accuracy of the estimator. It is shaped by the transfer function of the preamplifier topology. In this work, it will be considered a receiver with photodetector with PIN junction diode and field effect transistor (FET) transimpedance amplifier (TIA) \cite{receiver_design,2004_komine}.  The noise in the receiver is mainly composed by the shot noise and the thermal noise. Such noise sources can be modeled as Gaussian processes with zero mean and variances \cite{2012_ghassemlooy,2016_3-D_RSS,2004_komine}:
\begin{equation}
\sigma^2_n = \sigma^2_{\rm shot} + \sigma^2_{\rm thermal}\\
= \sigma^2_{\rm bg}  + \sigma^2_{\rm rs} + \sigma^2_{\rm dc} + \sigma^2_{\rm thermal}.
\label{eq:noise}
\end{equation} 

The photo-generated shot noise corresponds to the fluctuations in the count of the photons collected by the receiver \cite{2004_komine, 2014zhang}. The variances of the shot noise due to the background radiation ($_{\rm bg}$), the received signal ($_{\rm rs}$), and the dark current ($_{\rm dc}$) can be determined respectively by:
\begin{equation} 
\sigma _{\rm bg} ^2 = 2qR_{pd}A_{pd}p_{bs}\Delta\lambda B,
\label{eq:sigma2_bg}
\end{equation} 
\begin{equation} 
\sigma _{\rm rs} ^2 = 2qR_{pd}P_{R}B,
\label{eq:sigma2_rs}
\end{equation} 
\begin{equation}
\sigma _{\rm dc} ^2 = 2qI_{\rm dc}B,
\label{eq:sigma2_dc}
\end{equation} 
where $q$ is the elementary charge, $p_{bs}$ is the background spectral irradiance, $\Delta\lambda$ is the bandwidth of the optical filter, $B$ is the equivalent noise bandwidth and $I_{\rm dc}$ is the dark current.

The thermal noise is independent of the received optical signal and can be determined in terms of noise in the feedback resistor and noise in the FET channel. Each term, respectively, contributes to the following variance \cite{receiver_design,2004_komine}:
\small
\begin{equation} 
\sigma _{\rm thermal} ^2 = \frac{{8\pi k_B T_K }}{{G_{ol} }}C_{pd} A_{pd}I_2 B^2 + \frac{{16\pi ^2 k_B T_K \Gamma }}{{g_m }}C_{pd}^2 A_{pd}^2 I_3 B^3,
\label{eq:sigma2_thermal} 
\end{equation}
\normalsize 
where $k_B$ is the Boltzmann's constant, $T_K$ is the absolute temperature, $G_{ol}$ is the open loop gain, $C_{pd}$ is the capacitance per unit area of the photodetector, $\Gamma$ is the FET channel noise factor, $g_m$ is the FET transconductance, $I_2=0.562$ is the TIA bandwidth factor, and $I_3=0.0868$ is the TIA noise factor.

\subsection{RSS localization}\label{subsec:rss}
If the lighting infrastructure and distribution of the luminous flux of the LED transmitter are known, the receiver can determine its location by the luminous RSS information \cite{IEEEhowto:sahin}. Thus, to obtain a  minimum variance unbiased estimator (MVUE), the following observation vector can be considered:
\begin{equation}
\textbf{s} = \textbf{p}(\bm{\theta} ) + \textbf{n} \quad \in \mathbb{R}^{MK \times 1},
\label{eq:vetor_observacao}
\end{equation}
where $ \bm{\theta} \in {\mathbb{R}}^{3 \times 1}$ is the vector that corresponds to the exact location of the VLC receiver, {\it i.e.}, $\bm{r}_R$,  $\textbf{n} \in \mathbb{R}^{KM \times 1}  \sim \mathcal{N}(0,\sigma _n ^2 \textbf{I}_{KM} )$ is the additive noise vector. The vector  $ \textbf{p}(\bm{\theta} ) \in \mathbb{R}^{MK \times 1}$ is the vectorization of the matrix $\textbf{P}(\bm{\theta} )$. The matrix $\textbf{P}(\bm{\theta} )\in\mathbb{R}^{M \times K}$ contains the exact RSS information in the $m$-th  row referring to the $m$-th LED transmitter and the $k$-th column referring to the $k$-th VAP.

Considering the noise as additive white Gaussian noise (AWGN), the log-likelihood function (LLF) for the location of the VLC receiver can be expressed as:
\begin{equation}
\mathcal{L}(\bm{\theta} ) = \log (pdf(\bm{s},\bm{\theta} )),
\end{equation}
where the joint probability density function (PDF) is given by:
\begin{small}
	\begin{equation}
	pdf(\bm{s},\bm{\theta} ) = \frac{1}{({2\pi \sigma _n ^2 })^{\frac{MK}{2}}}\exp \left( { - \frac{1}{{2\sigma _n ^2 }}\left( {\bm{s} - \bm{p}(\bm{\theta} )} \right)^\intercal \left( {\bm{s} - \bm{p}(\bm{\theta} )} \right)} \right).
	\end{equation}
\end{small}

The joint PDF can be obtained from the product between the marginal PDFs due to the consideration they are independent and identically distributed. Applying the {$\log(.)$} operator, the maximum-likelihood estimation (ML) of $\bm{r}_R$ {problem} can be formulated by considering only the matrix operation of the exponential argument as:
\begin{equation}
\bm{\hat r}_R  = \arg \mathop {\max }\limits_{\bm{\theta}}  \mathcal{L}(\bm{\theta} )\, {\equiv} \arg \mathop {\max }\limits_{\bm{\theta}}   \left( { - \left( {\bm{s} - \bm{p}(\bm{\theta} )} \right)^\intercal \left( {\bm{s} - \bm{p}(\bm{\theta} )} \right)} \right).
\label{eq:ml}
\end{equation}
As a result, \eqref{eq:ml} can be expressed as a nonlinear least squares (NLLS) problem given by:
\begin{equation}
\bm{\hat r}_R  = \arg \mathop {\min }\limits_{\bm\theta}  \left( {\left\| {\bm{s} - \bm{p}(\bm{\theta} )} \right\|_2^2 } \right).
\label{eq:NLS_RSS}
\end{equation}

That way, this estimator minimizes the Euclidean distances between the observation vector $\bm{s}$ and the exact value of received intensities $\bm{p}(\bm{\theta}) $. Thus, one method to solve the system of nonlinear equations is that of Newton-Rapson Multivariate \cite{IEEEhowto:sahin,IEEEhowto:kay}:
\begin{equation}
\bm{\theta} ^{i + 1}  = \bm{\theta} ^i  - \eta \bm{J}^{\dagger}\left( \bm{s} - \bm{p}(\bm{\theta} ^i ) \right),
\label{eq:nlls}
\end{equation}
where $\eta \in (0,1]$ is the step size and $\bm{J}$ is the Jacobian matrix of $\bm{p}(\bm{\theta})$ in relation to $\bm{\theta}$.
Whereas $\theta_1$, $\theta_2$ and $\theta_3$ correspond to the positions $x$, $y$ and $z$ of VLC receiver. $\bm{J}$ can be given by:
\begin{equation}
\bm{J} = \left[ {\begin{array}{*{20}c}
	{\frac{{\partial P_{11} }}{{\partial x}}} & {\frac{{\partial P_{11} }}{{\partial y}}} & {\frac{{\partial P_{11} }}{{\partial z}}}  \\
	{\frac{{\partial P_{21} }}{{\partial x}}} & {\frac{{\partial P_{21} }}{{\partial y}}} & {\frac{{\partial P_{21} }}{{\partial z}}}  \\
	\vdots  &  \vdots  &  \vdots   \\
	{\frac{{\partial P_{MK} }}{{\partial x}}} & {\frac{{\partial P_{MK} }}{{\partial y}}} & {\frac{{\partial P_{MK} }}{{\partial z}}}  \\
	\end{array}} \right].
\label{eq:jacobian}
\end{equation}

In \eqref{eq:jacobian}, each row of $\bm{J}$ indicates how the RSS of each LED transmitter changes when the receiver moves on one of the axes, $x$, $y$ and $z$. Considering the chain rule ($ \frac{{\partial P}}{{\partial x}} = \frac{{\partial P}}{{\partial a}}\frac{{\partial a}}{{\partial x}}$), the row associated with the $m$-th LED transmitter of the $k$-th VAP can be calculated as:
\begin{equation}
\left[ {\begin{array}{*{20}c}
	{\frac{{\partial P_{mk} }}{{\partial x}}}  \\
	{\frac{{\partial P_{mk} }}{{\partial y}}}  \\
	{\frac{{\partial P_{mk} }}{{\partial z}}}  \\
	\end{array}} \right]^\intercal  = \left[ {\begin{array}{*{20}c}
	{\frac{{\partial P_{mk} }}{{\partial a_{mk} }}}  \\
	{\frac{{\partial P_{mk} }}{{\partial b_{mk} }}}  \\
	{\frac{{\partial P_{mk} }}{{\partial c_{mk} }}}  \\
	\end{array}} \right]^\intercal \left[ {\begin{array}{*{20}c}
	{\frac{{\partial a_{mk} }}{{\partial x}}} & {\frac{{\partial a_{mk} }}{{\partial y}}} & {\frac{{\partial a_{mk} }}{{\partial z}}}  \\
	{\frac{{\partial b_{mk} }}{{\partial x}}} & {\frac{{\partial b_{mk} }}{{\partial y}}} & {\frac{{\partial b_{mk} }}{{\partial z}}}  \\
	{\frac{{\partial c_{mk} }}{{\partial x}}} & {\frac{{\partial c_{mk} }}{{\partial y}}} & {\frac{{\partial c_{mk} }}{{\partial z}}}  \\
	\end{array}} \right].
\label{eq:linha_associada}
\end{equation}

From \eqref{eq:vmk}, the matrix in \eqref{eq:linha_associada}, {\it i.e.}, the Jacobian of $\bm{v}_{mk}$ with respect to $\bm{\theta}$, becomes an identity matrix. Therefore, \eqref{eq:linha_associada} can be directly obtained by evaluating the partial derivatives of \eqref{eq:power_light_received} by keeping one of the elements of the incidence vector $\bm{v}_{mk}$ as variable and the other elements as constants\footnote{{Notice that the proposed approach allows the use of a different light distribution than the Lambertian one, provided that its function is continuous and differentiable.}}. Thus, the partial derivative in relation to the first element of $\bm{v}_{mk}$ can be written as:
\begin{equation}
\begin{array}{l}
\frac{{df(\bm{v}_{mk} )}}{{da_{{{mk}}} }} = \frac{{\left( {\bm{v}_{mk} ^\intercal \bm{n}_{mk} } \right)^{\bm{n}_L } }}{{\left\| {\bm{v}_{mk} } \right\|_2 ^{{n}_L  + 3} }} \times \\ 
\left( {\begin{array}{*{20}c}
	{{n}_{{R}} ^{(x)}  + {n}_{mk} ^{(x)} \frac{{{n}_L \,\,\left( {\bm{v}_{mk} ^\intercal \bm{n}_R } \right)}}{{\,\left( {\bm{v}_{mk} ^\intercal \bm{n}_{mk} } \right)}} - a_{{{mk}}} \frac{{\,\left( {{n}_L  + 3} \right)\,\left( {\bm{v}_{mk} ^\intercal \bm{n}_R } \right)\,}}{{\left\| {\bm{v}_{mk} } \right\|_2^2 }}} 
	\end{array}} \right),
\end{array}
\end{equation}
being analogous for $\frac{{df(\bm{v}_{mk} )}}{{db_{{{mk}}} }}$ and $\frac{{df(\bm{v}_{mk} )}}{{dc_{{{mk}}} }}$ by changing the element $a_{mk}$ for $b_{mk}$ and $c_{mk}$, respectively, and changing the elements of $\bm{n}_{mk}$ and $\bm{n}_R$ in relation to $y$ and $z$, also respectively. Finally, the row vector can be determined by:
\small
\begin{equation}
\left[ {\begin{array}{*{20}c}
	{\frac{{\partial P_{mk} }}{{\partial x }}}  \\
	{\frac{{\partial P_{mk} }}{{\partial y }}}  \\
	{\frac{{\partial P_{mk} }}{{\partial z }}}  \\
	\end{array}} \right]^\intercal  =  \kappa \cdot P_T  \cdot  {\prod} \left( {\frac{{\theta _{mk} }}{{\theta _{FoV} }}} \right) {\prod} \left( {\frac{{\phi _{mk} }}{{\pi /2}}} \right) \left[ {\begin{array}{*{20}c}
	{\frac{{\partial f(\bm{v}_{mk}) }}{{\partial a_{mk} }}} \\
	{\frac{{\partial f(\bm{v}_{mk})}}{{\partial b_{mk} }}}  \\
	{\frac{{\partial f(\bm{v}_{mk}) }}{{\partial c_{mk} }}}  \\
	\end{array}} \right]^\intercal.
\label{eq:linha_assoc_completa}
\end{equation}
\normalsize

Although the Lambertian distribution model offers a convex set, the NLLS solution of \eqref{eq:NLS_RSS} is not convex in general \cite{IEEEhowto:sahin}. This is due to the fact that the set of possible solutions associated with each LED transmitter can become closer to each other at various locations in the 3-D geometry.  {This implies that the NLLS estimator can converge to one of the local optima.}
This inconvenience can be minimized by determining a better start-point for the NLLS estimator of \eqref{eq:nlls}. Hence, it can be utilized as an AoA estimator to determine the initial point $\bm{\theta}^0$.

\subsection{AoA Localization}\label{subsec:aoa}

In AoA localization method, the receiver checks and selects the transmitter LED that has the highest RSS for each VAP. Then, the estimator searches for a point that minimizes the sum of the distances (or quadratic distances) between the receiving point and between all other lines that extend in the direction of the $K$ LED transmitters selected by the receiver \cite{IEEEhowto:sahin}.

\begin{figure}[!htbp]
	\centering
	\includegraphics[width=.6\textwidth]{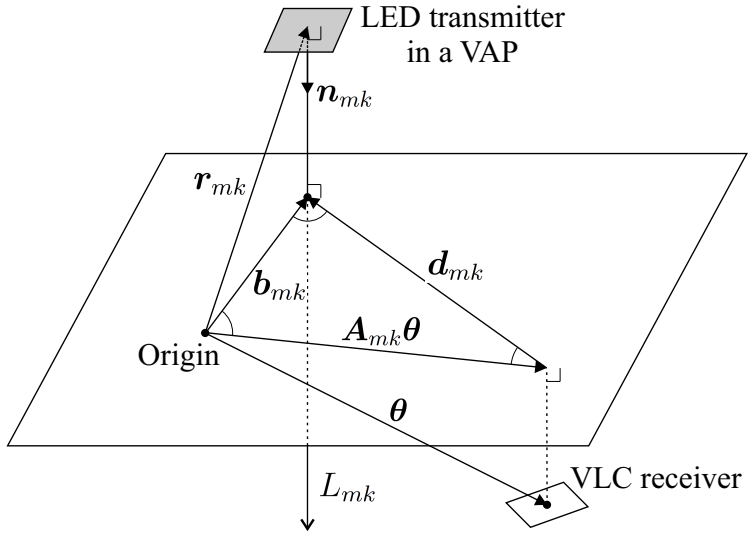}
	\caption{Geometry of AoA localization. Adapted from \cite{IEEEhowto:sahin}.}
	\label{fig:AOA_geometria}
\end{figure}

Fig. \ref{fig:AOA_geometria} depicts the geometry of the AoA location including the distances between the VLC receiver positioned in $\bm{\theta}$ and the direction defined by the line ${L}_{mk}$ extended in the normal direction of an LED transmitter positioned on the top of the room. That is,  ${L}_{mk}$ is the line that intersects $\bm{r}_{mk}$, collinear with $\bm{n}_{mk}$ and perpendicular to the plane. The matrix $\bm{A}_{mk} \in \mathbb{R}^{3 \times 3}$ projects any vector in the null space of $\bm{n}_{mk}$, that is, $ker\{\bm{n}_{mk}\}$, which can be calculated by:
\begin{equation}
\bm{A}_{mk}=\bm{I}_3-\bm{n}_{mk}\bm{n}_{mk}^\intercal.
\end{equation}

Consequently, the entire vector in the column space of $\bm{A}_{mk}$ is orthogonal to the direction of ${L}_{mk}$. The intersection point $\bm{b}_{mk}$ can be obtained by projection of $\bm{r}_{mk}$ vector in plane $ker\{\bm{n}_{mk}\}$, {\it i.e.}:
\begin{equation}
\bm{b}_{mk}=\bm{A}_{mk}\bm{r}_{mk},
\end{equation}
and the vectorial distance between the point $\bm{\theta}$ and the line ${L}_{mk}$ can be given by:
\begin{equation}
\bm{d}_{mk}=\bm{b}_{mk}-\bm{A}_{mk}\bm{\theta}.
\end{equation}

Stacking the set of equations related to the selected LED transmitters we have:
\begin{equation}
\bm{d} = \bm{b}-\bm{A}\bm{\theta}.
\label{eq:aoa_modelo_linear}
\end{equation}

As reported at the beginning of this section, the estimator searches for a point that minimizes the sum of the quadratic distances. In this way, the objective function can be established as:
\begin{equation}
\bm{\tilde r}_R  = \arg \mathop {\min }\limits_{\bm{\theta}}  \sum\limits_{\mathop {k = 1}\limits_{m = \xi_k } }^K {\left\| {\bm{d}_{mk} } \right\|_2^2 }  = \arg \mathop {\min }\limits_{\bm{\theta}}  \sum\limits_{\mathop {k = 1}\limits_{m = \xi_k } }^K {\left\| {\bm{A}_{mk} \bm{\theta}  - \bm{b}_{mk} } \right\|_2^2 } ,
\end{equation}
where $\xi_k$ is the index of LED transmitter that allowed the reception of greater luminous power from $k$-th VAP. Indeed, the AoA information considered is given by the LED transmitter of each VAP which is more directed to the receiver.

Considering that the concatenated matrix $\bm{A}$ is invertible, an estimator can be obtained by:
\begin{equation}
\bm{\tilde r}_R = \bm{A}^{\dagger} \bm{b}.
\label{eq:aoa_lse}
\end{equation}

However, identically treating the AoA information of each transmitter may cause a greater error in position estimation. This is because the AoA information of a farther transmitter LED has a lower signal-to-noise ratio (SNR) if compared to a closer transmitter LED. 
In other words, less reliable AoA information of the most distant LED transmitters causes a bias in location estimation and degrades location accuracy. To mitigate this problem, it is possible to consider an objective function that minimizes the weighted sum of the quadratic distances:

\begin{equation}
\bm{\hat r}_R = \arg \mathop {\min }\limits_{\bm{\theta}}  \sum\limits_{\mathop {k = 1}\limits_{m = \xi } }^K  {\beta _{mk} \left\| {\bm{A}_{mk} \bm{\theta}  - \bm{b}_{mk} } \right\|_2^2 } ,
\label{eq:aoa_lse_ponderado}
\end{equation}
where $\beta_{mk}$ is the weighting factor for the distance between $\bm{\theta}$ and $L_{mk}$.

In \cite{IEEEhowto:sahin}, it is analyzed the AoA of LED transmitters based on weighted RSS information directly, {\it i.e.}, weighted by the respective $P_{mk}$. Thus, the problem given in {\eqref{eq:aoa_lse_ponderado}} corresponds to an unconstrained quadratic optimization problem that can be solved by the LS method as:
\begin{equation}
\bm{\hat r}_R = {\bm{A}_W}^\dagger \bm{b}_W,
\end{equation}
where $\bm{A}_W  = \sum\limits_{\mathop {k = 1}\limits_{m = \xi } }^K {P_{mk} \bm{A}_{mk} }$ and $\bm{b}_W  = \sum\limits_{\mathop {k = 1}\limits_{m = \xi } }^K {P_{mk} \bm{b}_{mk} } $,

This weighted AoA (WAoA) provides a lower RMSE of the position if compared to the unweighted version \cite{IEEEhowto:sahin}. Because this estimator considers more RSS information obtained from the stronger signals, {\it i.e.}, with a higher signal to noise ratio (SNR) than those obtained by the weaker signals.

\section{Spatial Optical OFDM for power discrimination of multiple LEDs}\label{sec:multiple_leds}

\begin{figure*}[!t]
	\centering
	\includegraphics[width=1\textwidth]{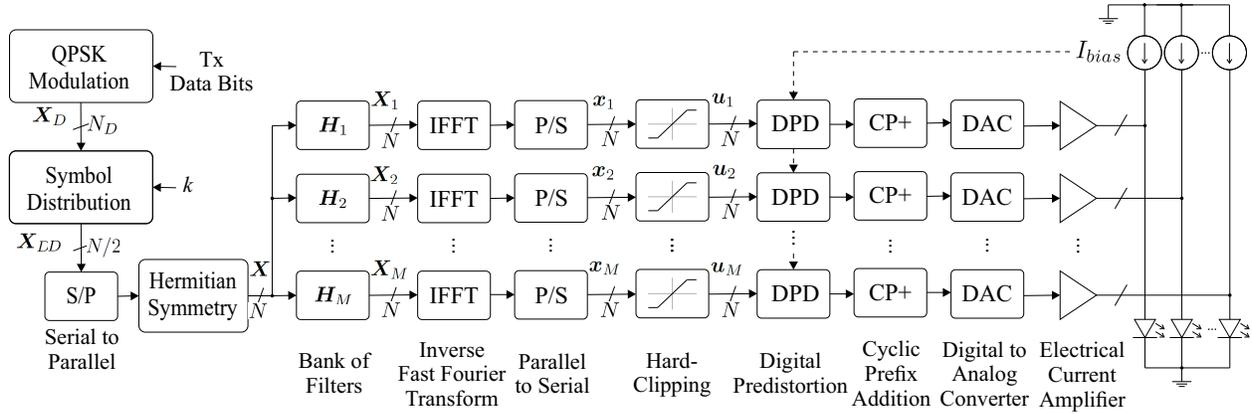}
	\caption{VLC-OFDM transmitter architecture implemented in each VAP.}
	\label{fig:arquitetura}
\end{figure*}

The localization estimators need to discriminate the powers. In this sense, the SO-OFDM \cite{vlc_multiple_leds} is deployed to divide the OFDM subcarriers emitted by each LED. Beyond reducing the PAPR, this scheme allows transmitting different signals on each of the $M$ transmitter LEDs in each VAP. Such strategy permits the signal discrimination in the receiver by applying the hybrid estimator discussed in the previous sections.  Next subsections elaborate on the sub-blocks associated to the proposed VLC-OFDM transmitter architecture, Fig. \ref{fig:arquitetura}. Such topology allows two operation modes, i.e., data communication and spatial signal localization at the receiver side.

\subsection{Subcarrier Allocation for Hybrid VLC Localization and Data Transmission}\label{sec:subc_alloc_communication}

In the proposed scheme, the data bits are QPSK modulated, generating the symbol vector $\bm{X}_D$. The fixed power level of the location subcarriers is intended to simplify the operation of the location estimators\footnote{The proposed architecture also allows data-carrying subcarriers to use higher-order modulation with different power levels. In such case, the clipping analysis should be updated.}. Thus, without loss of generality, the elements are admitted scaled such that:
\begin{equation}
\mathbb{E}\left[ {\left| {\bm{X}_{D}[i]} \right|^2 } \right] = 1; \,\,\, i=1,...,N_D.
\label{eq:esperanca_X}
\end{equation} 

For localization purpose, the elements of vector $\bm{X}_D$ first are distributed in the vector \footnote{Due to the pre-allocated power,  these symbols can be deployed for channel estimation and receiver synchronization purpose.}:
\begin{equation}
\bm{X}_{DD} \left[ {(m - 1)\frac{{N }}{2M} + k} \right] = \bm{X}_D [m - 1]\,\,\,;m = 1,...,M;
\label{eq:dist_subcarrier_loc}
\end{equation}
where $m$ represents the LED transmitter index of the VAP, $k$ is the index of the VAP, $M$ is the total number of LEDs transmitters in each VAP, $K$ the total number of VAPs and $N$ is the size of IFFT. The value of $k$ must be unique and must be predefined on each VAP of the infrastructure. In this way, each VAP will transmit $M$ symbols (in $M$ subcarriers) and will not transmit in $(M-1)K$, making them available to other VAPs. This allows the discrimination of the powers of each VAP to the location estimator in the VLC receiver. 

The distribution described by \eqref{eq:dist_subcarrier_loc} leaves a residue of $N/2-M(K+1)$ subcarriers available. This work proposes two operation modes for the system. The first one, termed location only mode (LOM); in this case, all the power of the available optical modulation signal is used to allow a better accuracy of the estimator. In the second operation mode, named location and communication mode (LCM), the power is distributed between the location and data transmission subcarriers; in this case, the greater transmission of data occurs in detriment of the location accuracy.

In LCM mode, strategically, it is proposed that these data-transmission subcarriers be distributed among the transmitter LEDs which allowed higher SNR at the receiver side. Multiple access can be obtained by redistributing the data subcarriers between the VAPs next to each VLC receiver.

Considering the case of a VAP with better signal to noise ratio (SNR) in the receiver. This VAP can distribute the residual symbols of $\bm{X}_D$ by:
\begin{equation}
\bm{X}_{DD} \left[ {(m - 1)\frac{{N }}{2M} + K+1,...,m\frac{{N }}{2M}} -1\right] = \bm{X}_D [i],
\end{equation}
with ${m = 1,...,M}$ and $i = m,...,\frac{N}{2}$\footnote{Herein, it is consided that there is no concomitant transmission of data subcarriers in the same indoor environment. This is to avoid co-channel interference.}. Thus, the vector $\bm{X}_{D}$ has size $N_D=\frac{N}{2}-M(K+1)$.
The elements of the vector $\bm{X}_{DD}$ that were not assigned by the two previous rules are accepted as nulls, maintaining the length $N_{DD}=\frac{N}{2}$. 

\begin{figure}[!t]
	\centering
	\includegraphics[width=.67\textwidth]{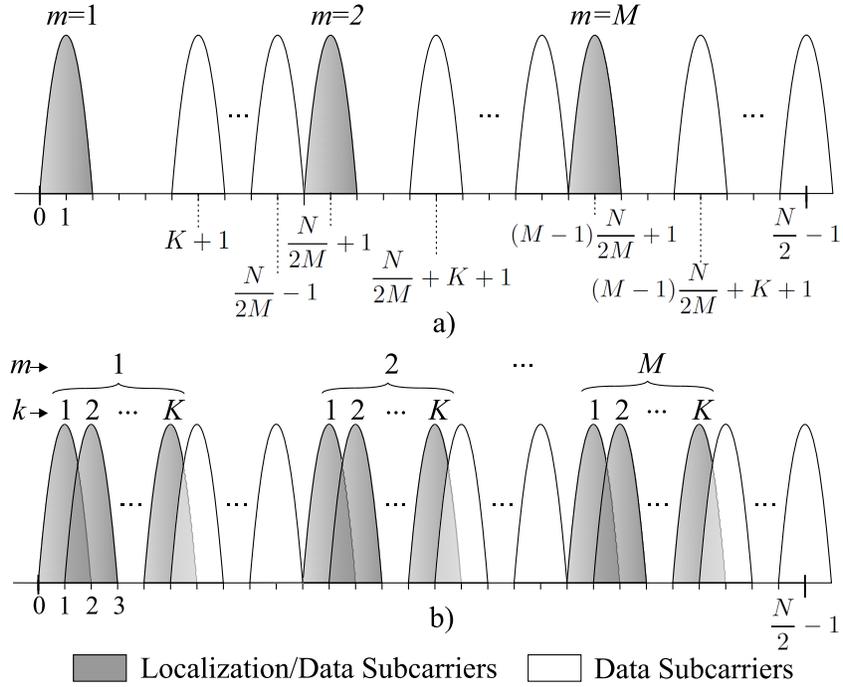}
	\caption{Proposed distribution of localization and data subcarriers for a) VAP transmitter $k=1$;  b) received subcarriers in VLC receiver.}
	\label{fig:distrib_subcarrier}
\end{figure}

Thus, in this symbol distribution scheme, the elements $\bm{X}_{DD}[0]$ and $\bm{X}_{DD}[\frac{N}{2}]$ responsible for the DC level are null; so, not interfering with the bias added to the OFDM frame in order to keep it positive. In this work, it is assumed that the control of this signal is performed externally to allow the control of the intensity of the illumination. Fig. \ref{fig:distrib_subcarrier} sketches the proposed subcarriers distribution in a VAP with $k=1$, as well as in a VLC receiver.

\subsection{DCO-OFDM}\label{sec:dco_ofdm}
In this work, the hybrid estimator is applied using the DCO-OFDM, although its application in the ACO-OFDM and Flip-OFDM should be equivalent.

After the serial to parallel conversion of $\bm{X}_{DD}$, with the purpose in obtaining purely real-time points at the out of the IFFT block, the vector $\bm {X}$ of size $N$ is generated after applying the following Hermitian symmetry:
\begin{equation}
\bm{X}[i] = \left\{ {\begin{array}{*{20}c}
	{\bm{X}_{DD} [i]} \hfill & {;i = 0, \ldots ,N/2-1 } \hfill  \\
	{\bm{X}_{DD} ^* [N - i]} \hfill & {;i = N/2, \ldots ,N - 1.} \hfill  \\
	\end{array}} \right.
\end{equation}

\subsection{SO-OFDM}\label{sec:so_ofdm}
The Hermitian vector $\bm{X}$ is then applied to the input of a filter bank of size $M$, {\it i.e.}, a filter for each $m$ LED of the VAP. Thus, the OFDM frames $\bm{X}_m$ are obtained by filtering the original OFDM frame $\bm{X}$ by means of:  
\begin{equation}
\bm{X}_m = \bm{H}_m \bm{X}.
\end{equation}

A contiguous mapping subcarrier with an equal number of subcarriers per LED is proposed by:     
\begin{equation}
\bm{H}_m[i]  = \left\{ {\begin{array}{*{20}c}
	{H,} \hfill & {i = (m - 1)\frac{{N }}{2M},...,m\frac{{N }}{2M}-1} \hfill  \\
	{\bf  0,} \hfill & \text{otherwise} \hfill  \\
	\end{array}} \right.
\label{eq:mat_H}
\end{equation}
where $H$ is a real-valued constant and $i = 0,1,...,\frac{N}{2}-1$. The second half of $\bm{H}_m$ is obtained by mirroring using $\bm{H}_m[i]=\bm{H}_m[N-1-i]$ for $i = \frac{N}{2},...,N-1$. In this way, the filters $\bm{H}_m$ masks the subcarriers that are not to be transmitted.

In each vector $ \bm{X}_m$ is performed the IFFT. Using the inverse discrete Fourier transform definition \cite{ofdm_owc} in the transmitter:
\begin{equation}
\bm{x}_m [i] = \frac{1}{{\sqrt N }}\sum\limits_{n = 0}^{N - 1} {\bm{X}_m [n] e^{j2\pi ni /N}} \,\, ;\,\textnormal{for} \,0 \le i \le N - 1;
\end{equation}
corresponds to a forward transform on the FFT block in the receiver by:
\begin{equation}
\bm{Y}[i] = \frac{1}{{\sqrt N }}\sum\limits_{n = 0}^{N - 1} {\bm{y}[n] e^{-j2\pi ni /N} } \quad ;\,\textnormal{for} \, 0 \le i \le N - 1;
\end{equation}
where $\bm{y}$ is the vector of the sample time domain signal and $\bm{Y}$ is the discrete frequency domain vector at the FFT output.

The signal in time obtained by IFFT block is converted from parallel to serial, obtaining the points in time $\bm{x}_m$ for each group. The signal is hard-clipped aiming at fit the dynamic range of the driver:
\begin{equation}
\bm{u}_m [i] = \left\{ {\begin{array}{*{20}c}
	{I_u } \hfill & {;\,\bm{x}_m [i] > I_u } \hfill  \\
	{\bm{x}_m [i]} \hfill & {;\,I_l  \le \bm{x}_m [ i] \le I_u } \hfill  \\
	{I_l } \hfill & {;\,\bm{x}_m [i] < I_l. } \hfill  \\
	\end{array}} \right.
\end{equation}

In the signal $\bm{u}_m$, the Digital Predistortion (DPD) is applied to correct the non-linearity associated with the direct current in the LED and the optical power obtained by it presented in {Subsection} \ref{subsec:system}. Also, the LED current of polarization of the LED must be considered in DPD step. Then the cyclic prefix (CP) is added, following digital to analog conversion (DAC), the current amplification, and finally the electric coupling to each LED of the VAP.

\subsection{Subcarrier Power Estimation in the VLC receiver}\label{sec:subc_power}

Given that the vector $\bm{x}_m$ is the sum of independent random variables with zero mean, using the central limit theorem (CLT) it is possible to approximate its Gaussian distribution of zero mean \cite{clipping_noise_ofdm_owc}. Considering also $N$ large enough, the variance of the $m$-{th} group of subcarriers can be given by $
\sigma ^2 _m  = \mathbb{E}\left[ {\left| {\bm{x}_m } \right|^2 } \right] $. Considering now \eqref{eq:esperanca_X} and \eqref{eq:mat_H} we obtain:
\begin{equation}
\sigma ^2 _m = \frac{2}{N}\sum\limits_{i = 0}^{N/2 - 1} {\left| {\bm{H}_m [i]} \right|^2 } = \frac{2}{N}\sum\limits_{i = 1}^{N_D /M} {H^2 }  = \frac{2}{N}\frac{{N_D }}{M}H^2.
\label{eq:sigma_m and N}
\end{equation}

The severity of clipping suffered by a signal is quantified by the clipping factor that is defined as the number of standard deviations per half of the dynamic range \cite{clipping_noise_ofdm_owc}:
\begin{equation}
\gamma _m  = { \frac{{I_u  - I_l }}{{2\cdot \sigma _m }}}.
\label{eq:Fator_Clipping}
\end{equation}

Considering the CLT approximation for $\bm{x}_m$, a symmetric clipping ($I_u=-I_l$), and the same variance $\sigma_m^2$ in all groups, the scaling factor $C{_\textsc{f} }$ can be determined as \cite{clipping_noise_ofdm_owc}:
\begin{equation}
C{_\textsc{f} } = 1 - \text{erfc}\left( {\frac{{\gamma _m }}{{\sqrt 2 }}} \right).
\label{eq:Fator_Escala}
\end{equation}

Applying \eqref{eq:Fator_Clipping} in \eqref{eq:Fator_Escala}, setting $I_u$ and $I_l$ and arbitrating the value of scaling factor $C{_\textsc{f}}$, the standard deviation $\sigma_m$ can be ready determined. In a same way, the constant value $H$ of the banks of filters $\bm{H}_m$  can be established using Eq. \eqref{eq:sigma_m and N}.  

Considering \eqref{eq:esperanca_X}, the magnitude of the $i$-th symbol at the receiver side can be expressed in the frequency domain by:
\begin{equation}
|\bm{Y}[i]|={H\cdot C{_\textsc{f} }\cdot G_E},
\label{eq:simbolo_recebido}
\end{equation}
where the electric gain $G_E$ is computed by \eqref{eq:electric_gain}.

\begin{figure}
	\centering
	\includegraphics[width=.5\textwidth]{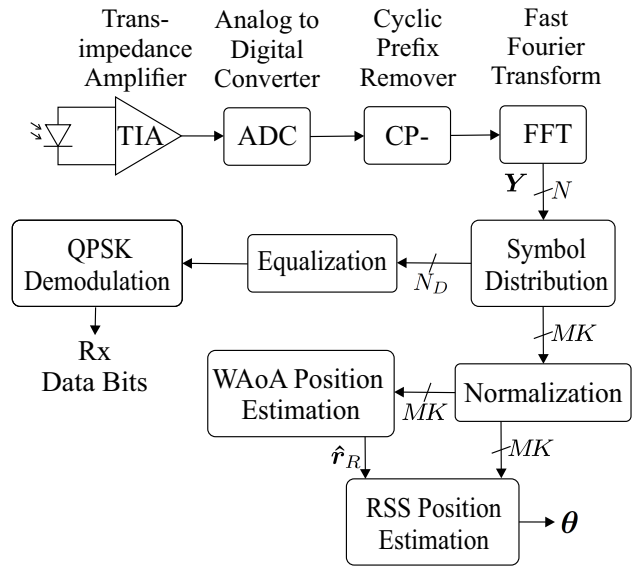}
	\caption{Architecture of proposed VLC receiver with receiver position estimation capability.}
	\label{fig:arquitetura_rx}
\end{figure}

After the separation of the magnitude of symbols destined for the location in the receiver, according to Fig. \ref{fig:arquitetura_rx}, the following normalization is performed in order to make the elements of the observed vector of RSS information in \eqref{eq:vetor_observacao} more suitable for the efficiency of the recursive location estimator:
\begin{equation}
s_{mk}=\frac{|{Y}_{mk}|}{HCR_{pd}S_{{\textsc{led}}}}={\Omega}_{mk}+n;
\label{eq:vetor_observado_eficiente}
\end{equation}
where $s_{mk}$, $Y_{mk}$ and $n$ are respectively the RSS information,  the magnitude of $i$-th symbol and the noise  sample in the receiver corresponding to the  $m$-th LED of the  $k$-th VAP.

Comparing  \eqref{eq:vetor_observado_eficiente} with \eqref{eq:power_light_received}, the transmitted power becomes unitary. Finally, it can be admitted the concatenation of the $s_{mk}$ elements in the vector of the exact RSS information $\bm{s}$ and as same for $\bm{p}(\bm{\theta})$ with the concatenation of $MK$ DC optic gain elements.

\begin{table*}[!h]
	\centering
	\caption{Adopted parameters values.}
	\begin{tabular}{l|l|l|l|l|l}
		\hline
		\multicolumn{1}{c|}{\bf Infrastructure}&\multicolumn{1}{c|}{\bf LED}&\multicolumn{1}{c|}{\bf Photodetector}&\multicolumn{1}{c|}{\bf Noise Model}&\multicolumn{1}{c|}{\bf OFDM}& \multicolumn{1}{c}{\bf Hybrid Estimator}\\
		\hline
		Room dim.: $5\times 4\times 3$ m & $n_L=10$    		& $A_{pd}=$ 1 cm$^2$ 			& $T_K=300$ K  		& $B=10$ MHz 	& $\bm{\theta}^0=\bm{\hat r}_R${, $\rm {C}$ or ${\rm RND}$} \\
		$K = 4$     					& $I_{\rm bias}=1.5$ A & $\theta_{\rm FoV}=85^o$ 	& $G_{ol}=10$   	& $N=1024$	 	& $\eta=$ 0.3 \\
		$M = 4$     					& $I_u=1$ A 		& $\bm{n}_R =[0, \, 0,\, 1]^\intercal$	& $g_m=30$ mS  		& $N_D=496$	 	& $\epsilon < 1\times10^{-5}\rm{m}$\\
		$\alpha = 20^o$ 				& $I_l=-1$ A 	& $R_{pd}=$ 0.54 A/W			& $\Gamma=1.5$		&  				& $i_{\max}=200$\\
		$\theta_{\rm ceiling} = 35^o$	&       			& $C_{pd}=$ 112 pF/cm$^2$  		& $\Delta\lambda$ = (780-380)nm=400 nm&   & \\
		$\theta_{\rm wall} = 45^o$ 		&       			& $I_{dc}=$ 5 pA	  			& $p_{bs}=5.8\mu$W/(cm$^2$nm)	&  	& \\
		\hline
	\end{tabular}%
	\label{tab:parameters}%
\end{table*}%

\section{Numerical Results}\label{sec:resultados}
In this section, we have demonstrated the effectiveness and efficiency of the proposed method by numerical simulation analysis. Similar to \cite{IEEEhowto:sahin}\footnote{Notice that this work also makes a numerical analysis about the best choices of the orientations and angles involved in the infrastructure components. Hence, the best values found for such parameters will be considered in our work for comparison purpose.}, the adopted infrastructure was an empty room with dimensions $5\times 4 \times3$ m where all four VAP devices are positioned in the four upper corners of the room ($K=4$). The VAPs' directions are $\theta_{\rm wall}=45^o$ between the walls and $\theta_{\rm ceiling}=35^o$ below the ceiling. Each VAP has four Cree\textsuperscript{\textregistered} XHP70.2 6V LED transmitters ($M=4$) in a pyramid shape with a square base as shown in Fig. \ref{fig:system_model2}a. The angle between the LED and the normal vector of each VAP is $\alpha = 15^o$. The VLC receptor parameters are $\theta_{\rm FoV} =85^o$, $A_{pd}=1\,\rm{cm}^2$ and $\bm{n}_R =[0, \, 0,\, 1]^\intercal$ \footnote{A priori knowledge of the receiver orientation is plausible since it can be estimated by a system composed by a three-axis accelerometers arrangement, commonly used in smartphones, among other portable devices.}. The noise parameters values deployed in our analyses are the same as that used in \cite{2004_komine}. In the RSS recursive estimator, the step value for the search algorithm was $ \eta = 0.3$. Initially, the  adopted stopping criterion for the recursive estimator was an error $\epsilon \leq 10^{-4} \rm{m}$ or a maximum number of iterations of the $i_{\max}=200$.

Such direct infrastructure parameter values, including parameters for the LED, photodetector, noise, the OFDM and for the recursive RSS estimator deployed in numerical analyses are summarized in the Table \ref{tab:parameters}. The indirect infrastructure parameter values obtained are the LED conversion factor $S_{\textsc{led}} =1.4812$ W/A applying eq. \eqref{eq:led_factor} and the noises variances $\sigma^2_{\rm bg}=4.0144\times 10^{-15}$ A$^2$, $\sigma^2_{\rm dc}=1.6022\times 10^{-23}$ A$^2$ and $\sigma^2_{\rm thermal}=6.5631\times 10^{-17}$ A$^2$ by applying \eqref{eq:sigma2_bg}, \eqref{eq:sigma2_dc} and \eqref{eq:sigma2_thermal}. The shot noise variance $\sigma^2_{\rm rs}$ was estimated for each analyzed position of the VLC receiver since it depends on the received light power from the transmitted signal.

Next, the analysis of the proposed architecture is divided into three parts. In these subsections, the performance, the clipping noise effect on the hybrid estimator performance, as well as the computational complexity of the estimators are evaluated.

\subsection{Performance of 3-D Localization Estimators}
\begin{figure}[htbp]
	\centering
	\vspace{-.3cm}
	\includegraphics[width=.7\textwidth]{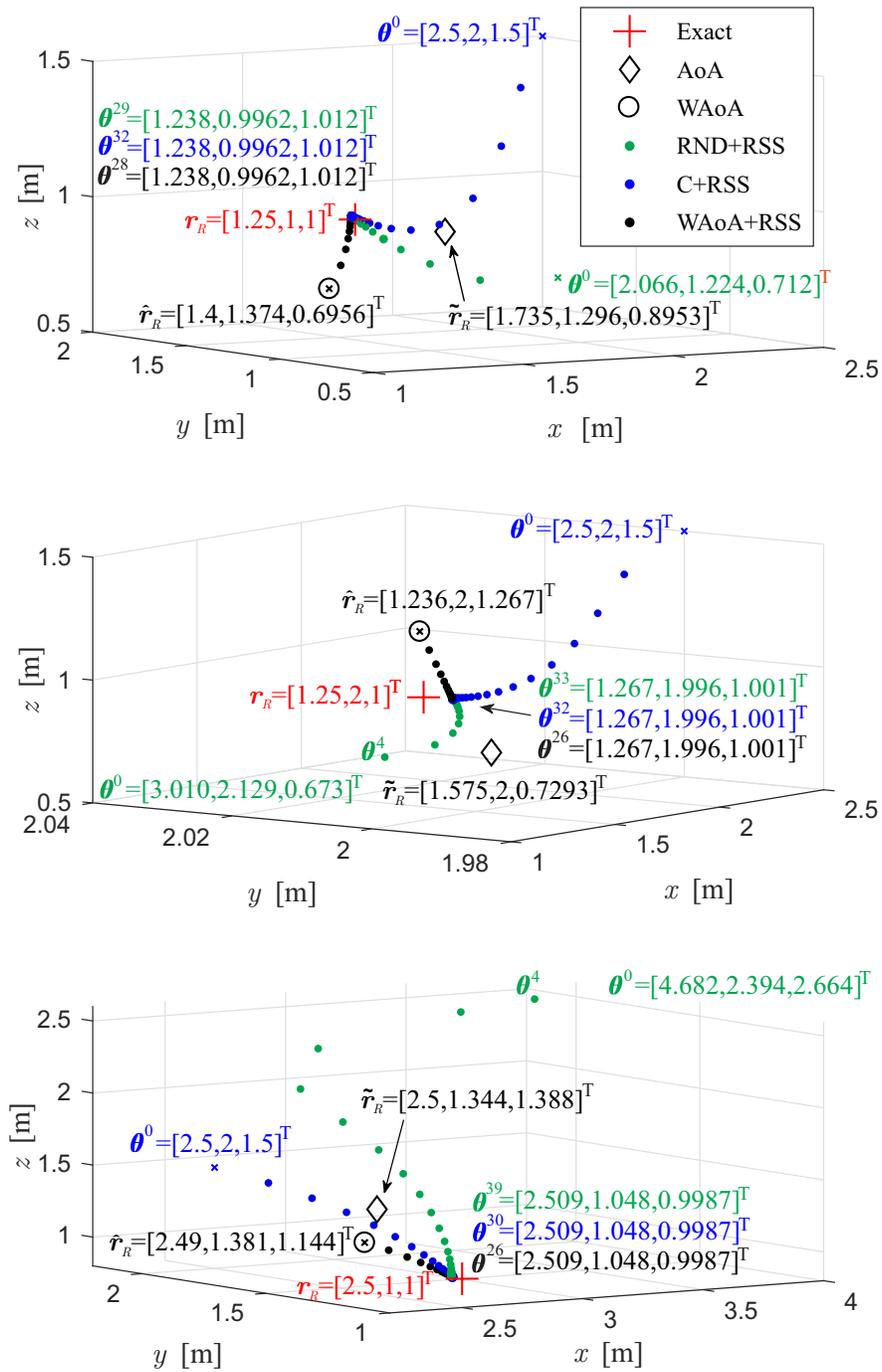}
	\caption{Localization convergence among the five 3-D localization methods for three different receiver positions given by the marker "$+$". The symbol $\times$ indicates the starting points $\bm{\theta}^0$ of the recursive estimators.}
	\label{fig:convergencia}
\end{figure}

{As a way of comparison, it has included the RSS estimator simulations considering the centroid of the room (C+RSS) and also a point obtained at random (RND+RSS) as starting points. In the latter case, the coordinates of the initial point $\bm{\theta}^0$ were obtained using a random variable with uniform distribution along the dimensions of the room. In the case evaluated $x\sim\mathcal{U}(0,5)$, $y\sim\mathcal{U}(0,4)$ and $z\sim\mathcal{U}(0,3)$. To compare the performance of the five localization methods, namely AoA, WAoA, C+RSS, RND+RSS and WAoA+RSS, the Euclidean error and number of iterations required for convergence were evaluated for one realization considering three different positions of the VLC receiver , a) $\bm{r}_{R,1}=[1.25,1,1]^\intercal$, b) $\bm{r}_{R,2}=[1.25,2,1]^\intercal$, and c) $\bm{r}_{R,3}=[2.5,1,1]^\intercal$ . 
	The numerical results of the system operating in LCM and also in LOM modes are shown in Table \ref{tab:convergencia}. All the estimators by RSS reached very close Euclidean errors, and these are much smaller than AoA and WAoA. As expected, in all RSS estimators, the LOM mode confirmed better results due to the greater emitted power of the localization signals. The Fig. \ref{fig:convergencia} depicts the graphical convergence behavior considering the five estimators final positions and the convergence of recursive estimators in LCM mode. The remarkable superiority of the WAoA+RSS hybrid method occurred both because of the smaller Euclidean error compared to AoA and WAoA, and because of the smaller number of iterations for the convergence of the result when compared to C+RSS and RND+RSS.}

\begin{table}[htbp]
	\centering
	\caption{Euclidean errors ({\bf a}) and number of iterations ({\bf b}) from the four localization estimators, considering three different positions and one realization.}
	\vspace{-.1cm}
	\begin{tabular}{|c|ccc|ccc|}
		\multicolumn{1}{c}{\rule{0pt}{3ex} ({\bf a}) }& \multicolumn{6}{c}{ $\left\| \bm{\tilde r}_R- \bm{r}_R \right\| _2 $ \, in \, [$mm$]}\\[1ex]
		\hline \rule{0pt}{3ex}
		mode		& \multicolumn{3}{c|}{\rule{0pt}{3ex}LCM} & \multicolumn{3}{c|}{LOM} \\
		\hline \rule{0pt}{3ex}
		VLC position & $\bm{r}_{R,1}$  & $\bm{r}_{R,2}$ 	& $\bm{r}_{R,3}$ 	& $\bm{r}_{R,1}$ 	& $\bm{r}_{R,2}$	& $\bm{r}_{R,3}$ \\
		\hline \rule{0pt}{3ex}
		AoA   	& 578.0	& 423.0	& 518.6	& 578.0 & 423.0	& 518.6 \\
		WAoA  	& 505.1	& 267.7 & 407.6	& 493.7	& 273.3	& 407.2 \\
		RND+RSS & 17.8	& 17.7	& 48.8	& 0.621 & 0.352 & 0.642 \\
		C+RSS	& 17.8	& 17.7	& 48.8	& 0.617 & 0.343 & 0.646 \\
		WAoA+RSS& 17.8	& 17.7	& 48.8	& 0.621 & 0.337 & 0.649 \\
		\hline
		\multicolumn{1}{c}{\rule{0pt}{3ex} ({\bf b})}& \multicolumn{6}{c}{\rule{0pt}{3ex} \bf Number of Iterations, $i$}\\
		\hline \rule{0pt}{3ex}
		mode	& \multicolumn{3}{c|}{\rule{0pt}{3ex}LCM} & \multicolumn{3}{c|}{LOM} \\
		\hline \rule{0pt}{3ex}
		VLC position  & $\bm{r}_{R,1}$  & $\bm{r}_{R,2}$ 	& $\bm{r}_{R,3}$ 	& $\bm{r}_{R,1}$ 	& $\bm{r}_{R,2}$	& $\bm{r}_{R,3}$ \\
		\hline \rule{0pt}{3ex}
		RND+RSS & 29    		& 33    		& 39    		& 29   		& 33    		& 41 \\
		C+RSS 	& 32    		& 32    		& 30   			& 32    	& 33    		& 33 \\
		WAoA+RSS& 28    		& 26    		& 26    		& 28   		& 27    		& 28 \\
		\hline
	\end{tabular}%
	\label{tab:convergencia}
\end{table}%
\vspace{-.1cm}

\begin{figure}
	\centering
	\includegraphics[width=.6\textwidth]{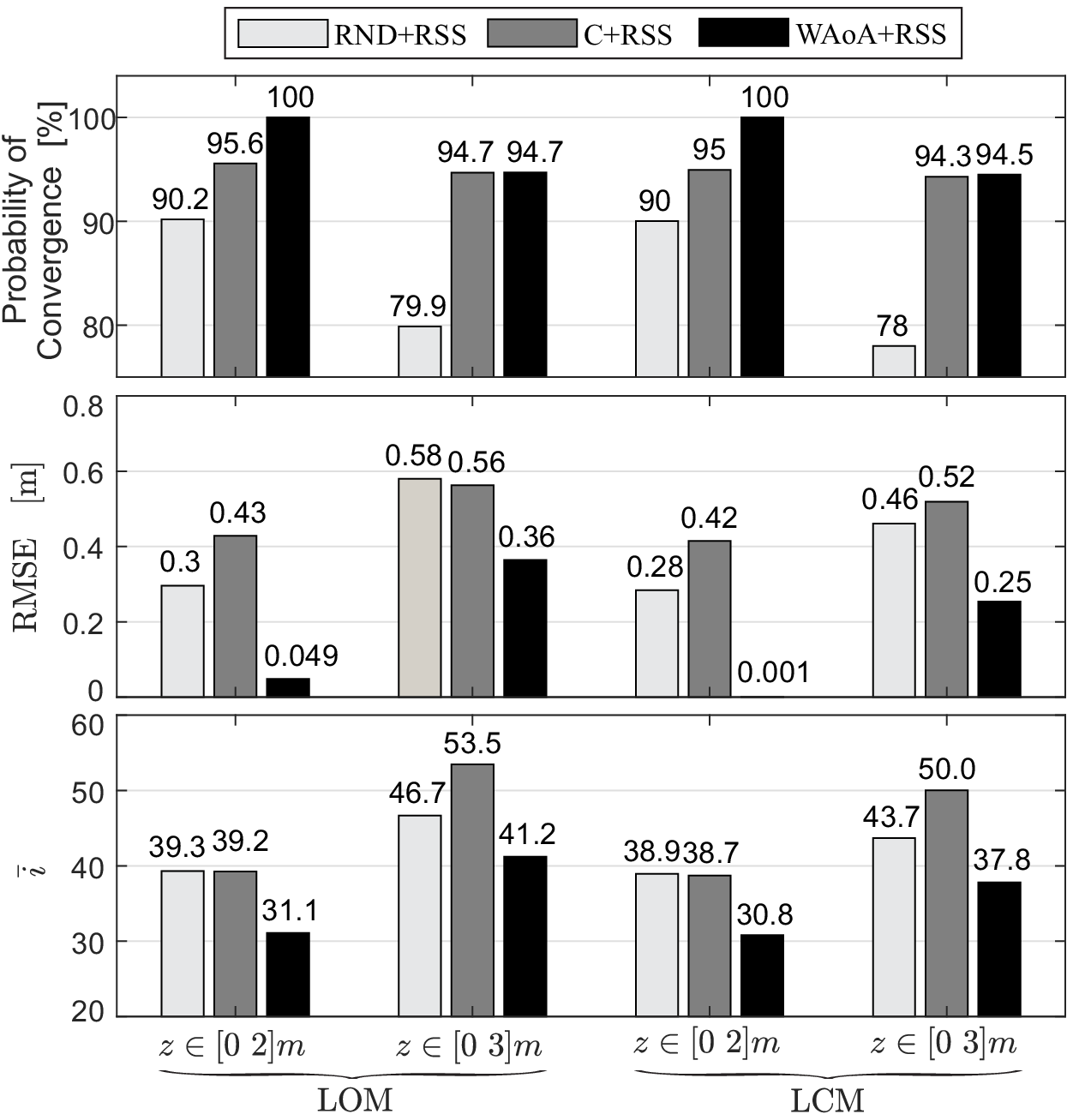}
	\caption{Statistics simulation of location estimators by RSS in the LCM and LOM modes of operation.}
	\label{fig:ConvRmseIter}
	\vspace{-.6cm}
\end{figure}

The Fig. \ref{fig:ConvRmseIter} shows the percentage of convergence, the RMSE and the average number of iterations $\bar i$ of the three analyzed RSS-based estimators for LCM and LOM operation modes. This analysis considered one hundred thousand achievements while keeping $\epsilon<1\times10^{-5}$ m. The exact positions of the VLC receiver were considered random with uniform distributions of probabilities of 
$\bm{r}_R  = \left[ {\begin{array}{*{20}c}
	{{\cal U}(0,5)} & {{\cal U}(0,4)} & {{\cal U}(0,2)}  \\
	\end{array}} \right]^\intercal$
and 
$\bm{r}_R  = \left[ {\begin{array}{*{20}c}
	{{\cal U}(0,5)} & {{\cal U}(0,4)} & {{\cal U}(0,3)}  \\
	\end{array}} \right]^\intercal$.
Therefore, in all receiver locations, the hybrid estimator confirmed its better performance by the higher percentage of convergence, smaller RMSE and smaller number of iterations if compared to the two others RSS-based estimators.

\begin{figure}[!htbp]
	\centering
	\includegraphics[width=1.02\textwidth]{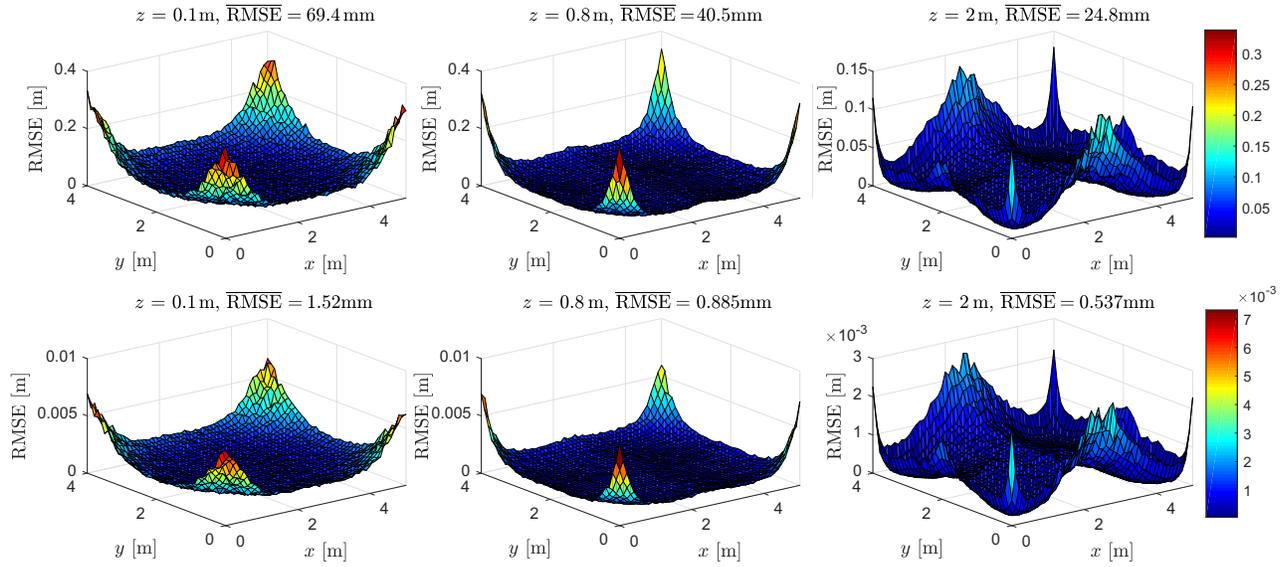}
	\caption{RMSE of the hybrid WAoA-RSS estimator over 100 realizations with the receiver located in  the three height planes, $z\in [0.1, \, 0.8,\, 2]^\intercal$ m. The first line of graphics refers to the LCM mode and the second to the LOM mode.}
	\label{fig:superficie_de_erro}
\end{figure}

To gain insight on the proposed hybrid WAoA-RSS performance, Fig. \ref{fig:superficie_de_erro} depicts the mean of 100 achievements of the RMSE of the proposed OFDM hybrid WAoA+RSS estimator for $z=0.1$, $z=0.8$  and $z=2\,\rm{m}$ height planes in the room. The achieved average $\rm{RMSE}$ value, i.e. $\overline{\rm RMSE}$, is shown for each graph.  Also, such numerical results considered a fixed quantity of $i=30$ iterations with pitch distance of analysis of $10$ cm. In general, the greater accuracy of the hybrid estimator is in the central region of the room where the luminous power is higher than in the corners. Indeed, for such location application the estimation of the {\it corners} reached an RMSE less than 350 mm (LCM mode) and 8 mm (LOM mode), while in the {\it central regions} it reached RMSE of less than 50mm (LCM mode) and 2 mm (LOM mode). In this way, the numerical results obtained under LOM mode have resulted in better localization accuracy than those presented in \cite{2015_3D_OFDM}. This result confirms the effectiveness and accuracy of the proposed method in a 3-D target object localization, representedd by  the optical receiver.

\begin{figure}[!htbp]
	\centering
	\vspace{-.5mm}
	\includegraphics[width=.65\textwidth]{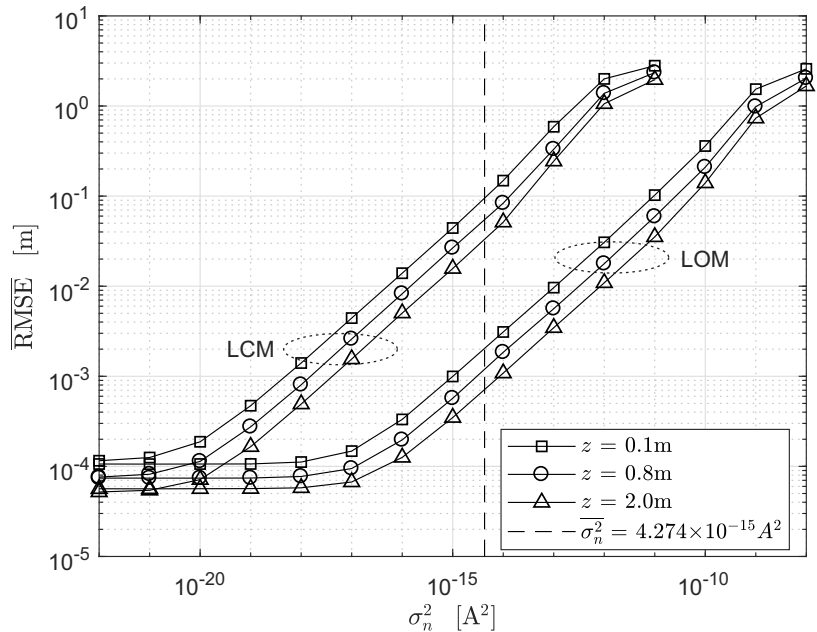}
		\vspace{-.5mm}
	\caption{$\overline{\rm RMSE}$ as function of fixed power of noise for the three height planes analyzed in Fig. \ref{fig:superficie_de_erro}.}
	\label{fig:rmse_noise}
\end{figure}

The variable portion of the noise power $\sigma_{\rm rs}^2$ contributed little to the total noise variance. In the simulated system environment, $\sigma_{\rm rs}^2$ noise was about one hundred times smaller than $\sigma_{\rm bg}^2$. Therefore, the location accuracy as a function of the fixed $\sigma^2_n$ was analyzed. Hence, the $\overline{\rm RMSE}$ over 100 realizations was obtained considering the same planes analyzed in Fig. \ref{fig:superficie_de_erro} and varying the $\sigma^2_n$ between $10^{-20}A^2$ and $10^{-8}A^2$. According to Fig. \ref{fig:rmse_noise}, it is observed that the RMSE decays a decade for each two decades of decrement in $\sigma^2_n$.

\subsection{Clipping Noise Effect}
The clipping generates a noise power that is added to the noise in the photodetector. The noise clipping variance can be estimated by \cite{2012clipping,dimitrov2015principles}:
\begin{equation}
\begin{array}{l}
\sigma _{\rm clip} ^2  = \sigma _m ^2 \left[ {C - C^2  + \left( {1 - Q(\lambda _l )} \right)\lambda _l ^2  + Q(\lambda _u )\lambda _u ^2 } \right. \\ 
- \left( {\varphi (\lambda _l ) - \varphi (\lambda _u ) + \left( {1 - Q(\lambda _l )} \right)\lambda _l  + Q(\lambda _u )\lambda _u } \right)^2  \\ 
\left. { + \varphi (\lambda _l )\lambda _l  - \varphi (\lambda _u )\lambda _u } \right], \\ 
\end{array}
\label{eq:sigma_clip}
\end{equation}
where $\lambda_l=I_l/\sigma_m$, $\lambda_u=I_u/\sigma_m$, with $\sigma_m=I_u/\gamma_m$ by \eqref{eq:Fator_Clipping}, 
$Q(v)$ is the Q-function, and the Gaussian function
\begin{equation}
\varphi (v) = \frac{1}{{\sqrt {2\pi } }}\exp \left( { - \frac{{v^2 }}{2}} \right).
\end{equation}

With a symmetrical clipping, i.e, $I_u=-I_l$, we have $\lambda_u=-\lambda_l$. Considering that $\varphi (v)$ is a even function and that the Q-function has the property $Q(v)=1-Q(-v)$, eq. \eqref{eq:sigma_clip} can be simplified:
\begin{small}
\begin{equation}
\hspace{-.1cm}
\sigma _{\rm clip} ^2  = \frac{I_u^2}{\gamma_m^2} \left[ C - C^2  + 2Q \left( {I_u}/{\gamma_m}\right)\frac{I_u^2}{\gamma_m^2} -2\varphi \left( {I_u}/{\gamma_m}\right)\frac{I_u}{\gamma_m} \right].
\end{equation}
\end{small}

Thus, the channel capacity with a symmetrical clipping for the $m$-th group of subcarriers can be determined by:
\begin{equation}
\mathfrak{C}_m  = \frac{B}{N}\frac{N_D}{M}\log _2 \left( {1 + \frac{{{I_u}^2/({2\gamma_m}^2)C G_E}}{{\left( {\sigma _n ^2  + \sigma _{\rm clip} ^2 } \right)\frac{{N_D }}{{NM}}}}} \right).
\label{eq:capacidade}
\end{equation}

\begin{figure}[htbp]
	\centering
	\includegraphics[width=.67\textwidth]{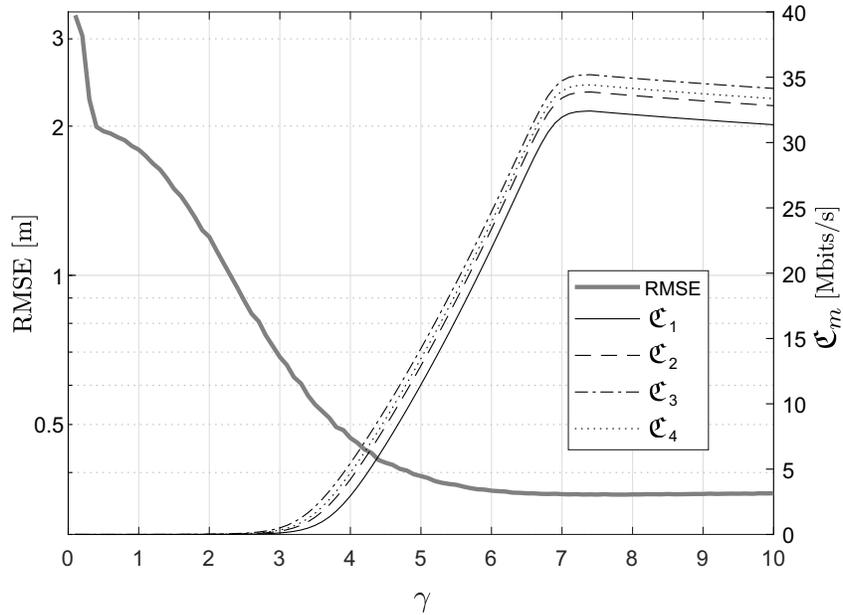}
	\caption{RMSE of locator and channel capacity of each $m$ LED in the $k=1$ VAP.}
	\label{fig:clipping_1position}
	\vspace{-.2cm}
\end{figure}

Fig. \ref{fig:clipping_1position} portrays an evaluation of the RMSE of the locator in LCM mode obtained as a function of the clipping factor $\gamma$ considering an exact location of the receiver as $\bm{r}_{R,1}=[1.25,\, 1, \, 1]^\intercal$ with 1000 realizations. Fig. \ref{fig:clipping_1position} also presents the theoretical capacity limit of data transmission for the $m$-th LED of the VAP $k = 1$ with better SNR at the receiver. It is observed that with a low clipping factor occurs severe degradation both in the localization estimates provided by the hybrid WAoA+RSS estimator and in capacity of data transmission. All capacity curves presented their optimal point for $\gamma_m \approx 7.4$. In this condition, the RMSE of the location estimation stabilized at the minimum level, confirming the best operating point of the system. Indeed, for $\gamma >7.4$, the channel capacity decreases very slightly due to the fact that a more significant $\gamma$ implies in a smaller $\sigma_m$ and consequently a lower SNR.  In this way, the optimal operation point for the location mode of the proposed architecture can be determined. Thus, considering the dynamic range of the LED and assuming the optimal clipping factor, the variance of the group of subcarriers $\sigma_m^2$ can be estimated by (\ref{eq:Fator_Clipping}) and then the constant $H$ of the filter banks by (\ref{eq:sigma_m and N}).

\subsection{Complexity of the Hybrid Localization Algorithm}
The number of multiplications and divisions was estimated for the computational complexity analysis of the estimators\footnote{It was considered the Gauss-Jordan algorithm for matrix multiplication and inversion of matrices \cite{1996van_loan}.}. Considering the pseudo-random generation using linear congruent generators \cite{1992numerical_recipes}, it would take about a dozen multiplications to obtain the initial search position in the RND+RSS estimator. In the AoA and WAoA estimators, the more computational resources-consuming matrix operation is the Moore-Penrose pseudo-inverse. For AoA and WAoA locator methods, the matrices $\bm{A}$ and $\bm{A}_w$ having dimensions $3K \times 3$ results in a pseudo-inverse complexity of $54K + 27$. Considering the multiplication by vector $\bm{b}$, the final complexity of AoA locator results in $\mathcal{C}_{\rm AoA}=63K + 27$. While for the WAoA, it requires $\mathcal{C}_{\rm WAoA}=75K + 27$ due to the weighting multiplications needed for the determination of the vector $\bm{b}_w$ and the matrix $\bm{A}_w$. Moreover, for the RSS locator, each determination of the Jacobian matrix $\bm{J}$ has complexity $MK(3n_L+48)/2$ and each iteration of the estimator requires $24MK + 27$.  As a result, a total of $\mathcal{C}_{\rm RSS}=MK(3n_L+99)/2+27$ operation per iteration is need.

\begin{figure}[htbp]
	\centering
	\includegraphics[width=.64\textwidth]{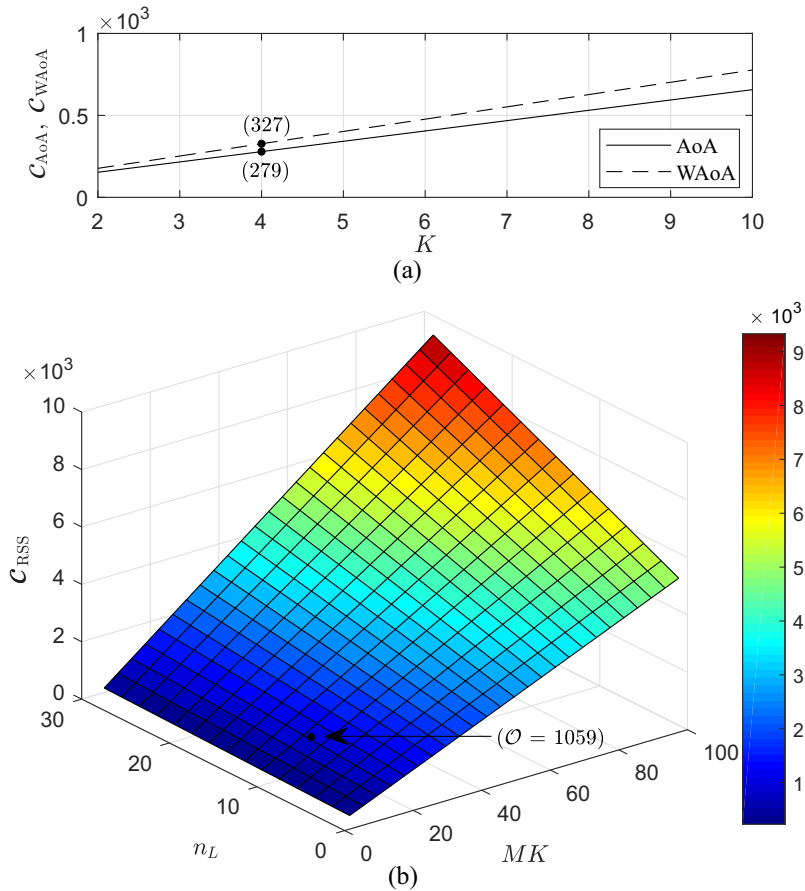}
	\caption{Complexity for the three locator methods: a) AoA, WAoA; and b) RSS. Markers "\textbullet" identify the complexities for the  previously analyzed configuration with parameters in Table \ref{tab:parameters}.}
	\label{fig:complexidade}
\end{figure}

Fig. \ref{fig:complexidade} depicts the complexity of AoA and WAoA locators in relation to $K$. In the same figure, it is presented the complexity surface for one iteration of RSS estimator regarding the product $M \cdot K$ and Lambertian mode $n_L$. The complexity of the RSS estimator grows linearly with both $M$, $K$ and $n_L$. Moreover, in all plots, the complexities for the analyzed system configuration with parameters of Table \ref{tab:parameters}  are also identified. Indeed, for this specific scenario, the complexity of one iteration of RSS estimator was roughly three times greater than the AoA, WAoA complexities. 

Considering that for lighting purpose an LED with high directivity is not attractive, the Lambertian distribution mode number $n_L$ should reach the value of a few tens \cite{vlc_multiple_leds}.  Thus, an asymptotic complexity of $\mathcal{O}(MK)$ for RSS locator may be allowed, being reasonable against the asymptotic complexities of $\mathcal{O}(K)$ for the AoA and WAoA locators. Hence, the complexity penalty of using the WAoA+RSS estimator can be accepted taking into account the reasonable accuracy achieved and a higher percentage of convergence.

\section{Conclusions}\label{sec:conclusion}
With the proposed SO-OFDM architecture it was possible to discriminate the transmitted power from each LED of each VAP at the photodiode receiver. The subcarrier power estimation referring to each LED at the receiver enabled the operation of the proposed hybrid estimator based on WAoA combined with the RSS method. Hence, our numerical results have demonstrated that the RSS-based recursive estimator becomes much more accurate with a manageable complexity increasing if the WAoA estimator provides the start point search. The WAoA locator, although results in a limited precision, it is close enough to allow convergence to the exact position within a few numbers of iterations. {The ability of the proposed system to attain better accuracy using the location-only mode (LOM) enabled an improvement of about 40 times compared to the integrated location and communication mode  (LCM) of operation. Specifically, for the plane of height $h = 0.8$ m analyzed, the mean RMSE was reduced from $\overline{\rm RMSE}_{\rm LCM}=40.5\,mm$ to  $\overline{\rm RMSE}_{LOM}= 0.885\,mm$. As a result, a greater  system flexibility is obtained, increasing the range of applications while maintaining ambient illumination functionally and allowing the simultaneous operation of the location estimator and data transmission.}

The numerical results also demonstrated the RMSE decays a decade for every two decades of decrement of noise power. The clipping noise analysis allowed to determine the optimum point of the system in terms of data transmission capacity and lower RMSE of the WAoA+RSS estimator.
{In this case, the sum of the maximum transmission capacity in the simulated scenario allowed data rates of 130 Mbits/s for analyzed VAP.}
 Hence, this parameter serves as a guide for the clipping limit determination that minimizes PAPR, allowing an improvement in the performance of the data VLC-OFDM system.


\section*{Acknowledgment}
This work has been partially supported by the National Council for Scientific and Technological Development (CNPq) of Brazil under Grants 304066/2015-0; by the Londrina State University (UEL) and the Paran\'a State Government. All the agencies are gratefully acknowledged.


\newpage
\end{document}